\documentclass[aps,showpacs,superscriptaddress,preprint]{revtex4}%
\usepackage{mathrsfs}
\usepackage{amsfonts}
\usepackage{amsmath}
\usepackage{amssymb}
\usepackage{graphicx}%
\providecommand{\U}[1]{\protect\rule{.1in}{.1in}}

\begin{document}
\title{Band structure and charge doping effects of potassium-adsorbed FeSe/SrTiO$_3$ system}
\author{Fawei Zheng}
\affiliation{Institute of Applied Physics and Computational Mathematics, Beijing 100088, China}
\author{Li-Li Wang}
\affiliation{Department of Physics, Tsinghua University, Beijing 100084, China}
\author{Qi-Kun Xue}
\affiliation{Department of Physics, Tsinghua University, Beijing 100084, China}
\author{Ping Zhang}
\thanks{Corresponding author: zhang\_ping@iapcm.ac.cn}
\affiliation{Institute of Applied Physics and Computational Mathematics, Beijing 100088, China}

\begin{abstract}
We theoretically study, through combining the density functional
theory and an unfolding technique, the electronic band structure and
the charge doping effects for the deposition of potassium (K) on
multilayer FeSe films grown on SrTiO$_3$ (001) surface. These
results form a theoretical base line for further detailed studies of
low-temperature electronic properties and their multiway quantum
engineering of FeSe thin films. We explain the Fermi surface
topology observed in experiment and formulate the amount of doped
electrons as a function of atomic K coverage. We show that the
atomic K deposition efficiently dopes electrons to top layer FeSe.
Both checkerboard and pair-checkerboard antiferromagnetic (AFM) FeSe layers
show electron pockets at M point and no Fermi pocket at $\Gamma$
point with moderate atomic K coverage.
The electron transfer from K adsorbate to FeSe film introduces a
strong electric field, which leads to a double-Weyl cone structure
at M point in the Brillouin zone of checkerboard-AFM FeSe. We demonstrate that with
experimentally accessible heavy electron doping, an electron-like Fermi pocket will emerge at
$\Gamma$ point, which should manifest itself in modulating the
high-temperature superconductivity of FeSe thin films.
\end{abstract}
\maketitle

The superconductivity in monolayer FeSe film grown on SrTiO$_3$
(001) surface has recently been observed
\cite{wang2012,liu2012-2,tan2013,he2014,peng2014-2,peng2014,cui2015,fan2015,huang2015}.
It is attracting extensive studies on this unique system
\cite{liu2012,bazhirov2013,yuan2013,berlijn2014,hao2014,cao2015,
liu2015,mazin2015,xie2015,yang2015,yi2015,zhao2015}. The monolayer
FeSe film is the fundamental building block of crystalline
iron-based superconductors. The FeSe/SrTiO$_3$ system provides us
with a nice platform to study the monolayer FeSe film thoroughly,
and may shed light on superconducting mechanism of the entire
iron-based superconductors. The superconducting transition
temperature T$_c$ of FeSe/SrTiO$_3$ exceeds that of bulk FeSe by one
order of magnitude \cite{wang2012,ge2015, zhang2015,hsu2008}.
Whereas, further grown FeSe layers on FeSe/SrTiO$_3$
\cite{liu2014,deng2014,wang2015} and FeSe layers on graphene
\cite{song2011} and CaF$_2$ \cite{nabeshima2013} do not show such a
high-T$_c$ enhancement, which indicates a critical role played by
the interaction between the monolayer FeSe and SrTiO$_3$ substrate.
This interface-induced T$_c$ enhancement is supposed to be closely
associated with the electron doping from SrTiO$_3$ substrate
\cite{zhang2014,shanavas2015,he2013,bang2013,zheng2013}, as well as
with the strength of electron-phonon interaction
\cite{wang2012,xiang2012,li2014,lee2014,coh2015,rademaker2015}.
Angle resolved photoemission spectroscopy (ARPES) measurements have
found that there are only electron-like Fermi pockets that locate at
M point, while the hole-like Fermi pocket at $\Gamma$ point
disappears \cite{liu2012-2,he2013,tan2013}. This peculiar Fermi
surface topology indicates that the system is strongly electron
doped. The amount of doped electrons is increased with increasing
annealing time, and finally reaches a saturated value. Recently, by
depositing potassium (K) atoms onto multilayer FeSe films grown on
SrTiO$_3$ surface (we henceforth name the system as
K/n-FeSe/SrTiO$_3$, where n is the number of FeSe layers), Miyata
{\it et al.} demonstrated that atomic K deposition can dope
electrons to FeSe films more effectively \cite{miyata2015}. In this
way, the electron doping can be tuned continually and reversely,
thus expanding the accessible doping range towards the heavily
electron-doped region. The superconducting phase diagram shows a
clear dome-like shape and the maximum T$_c$ is as high as 48 Kelvin
\cite{miyata2015}. Along this line, further experimental measurements are being carried out \cite{Feng2015,Xue2015}.  

\begin{figure}[ptb]
\begin{center}
\includegraphics[width=0.4\linewidth]{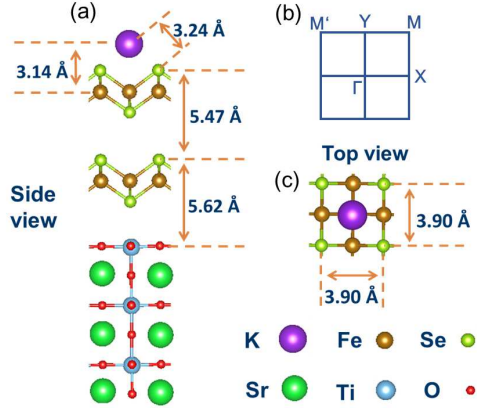}
\end{center}
\caption{(Color online) The side (a) and top (c) view of
K/2-FeSe/STO atomic structure. Panel (b) shows the two-dimensional
first Brillouin zone of K/n-FeSe/STO.}
\label{Fig1}%
\end{figure}

To self-consistently depict the above-mentioned charge doping and
predict new adsorbate-tuned electronic properties superposed onto
the inherently profound physics of the FeSe/SrTiO$_3$ system,
apparently, a systematic and enlightening theoretical study of K
deposition affection to the electronic structures of FeSe films is
essential but still keeps to be initialized. In this paper, by using
the density functional theory (DFT) based first-principles
calculations and checkerboard (CB) as well as pair-checkerboard (PCB) antiferromagnetic (AFM) models, the
energy band structure and the charge transfer effects of the
K/n-FeSe/SrTiO$_3$ system is studied. The Fermi surface topology
observed in experiment is satisfactorily explained. We find that
only the top layer FeSe is effectively charged. The amount of doped
electrons linearly depends on atomic K coverage (denoted by
$\rho_K$). The induced electric field strongly distorts the CB FeSe
energy bands, leading to a double-Weyl cone structure at M point.
Remarkably, our results also show that an electron-like Fermi pocket
will rapidly develop at $\Gamma$ point if the amount of doped
electrons further increases to be larger than 0.2 $e$ per FeSe
formula unit in CB FeSe. For PCB FeSe, the critical value is 0.17 $e$.

The DFT calculations in this work were mainly performed by using
{\it Vienna Ab-initio Software Package} (VASP) \cite{vasp}. The atom
core electrons were described by projector augmented wave (PAW)
method \cite{paw1,paw2}. Perdew-Burke-Ernzerhof \cite{pbe1} (PBE)
functional was used to treat the electronic exchange-correlation.
The energy cutoff for the plane-wave basis was set at 400 eV. The
first Brillouin zone was sampled in the $k$-space with
Monkhorst-Pack scheme and grid sizes are 13$\times$13, 7$\times$7,
 5$\times$5 and 8$\times$4 for 1$\times$1, 2$\times$2, 3$\times$3 and 2$\times$4 supercells,
respectively. We have checked that the total energy is converged for
the cutoff energy and $k$-point sampling. The atomic structure was
relaxed until the force on each atom is smaller than 0.01 eV/{\AA}. The
K/n-FeSe/SrTiO$_3$ can be divided into three parts (K atoms, FeSe
layers and SrTiO$_3$ substrate). Considering that weak attractions
exist in this composite system, we have added the van der Waals
correction \cite{vdw} to DFT calculations. The slab models were used
and more than 12 {\AA} vacuum space was added to avoid artificial
interaction. The in-plane lattice parameter was set to 3.901 {\AA}
\cite{zheng2013}. We used 6-layer SrTiO$_3$ with the TiO$_2$ layer
on the top to mimic the SrTiO$_3$ substrate. Besides the VASP code,
we also performed DFT calculations in {\it Quantum Espresso} (QE)
\cite{qe} code with norm-conserving pseudopotentials \cite{nc} to get
accurate unfolded electronic energy bands by using a home-built
code.

The relaxed atomic structure of the K/2-FeSe/SrTiO$_3$  is shown in
Figs. \ref{Fig1}(a,c). The $\rho_K$ is 1, which is one K atom per
FeSe primary cell. The thickness of the down layer FeSe, which is
defined by the distance between the top Se atom in FeSe and top
layer TiO$_2$, is 5.62 {\AA}. The thickness of the top layer FeSe is
5.47 {\AA}, which is very close to the $c$ lattice parameter of bulk
FeSe \cite{hsu2008}. The K atoms adsorb on top layer FeSe, right on
the top of bottom Se atoms. The distances between the K atom and top
Se atoms and Fe plane are determined to be 3.24 {\AA} and 3.14
{\AA}, respectively.

\begin{figure}[ptb]
\begin{center}
\includegraphics[width=0.7\linewidth]{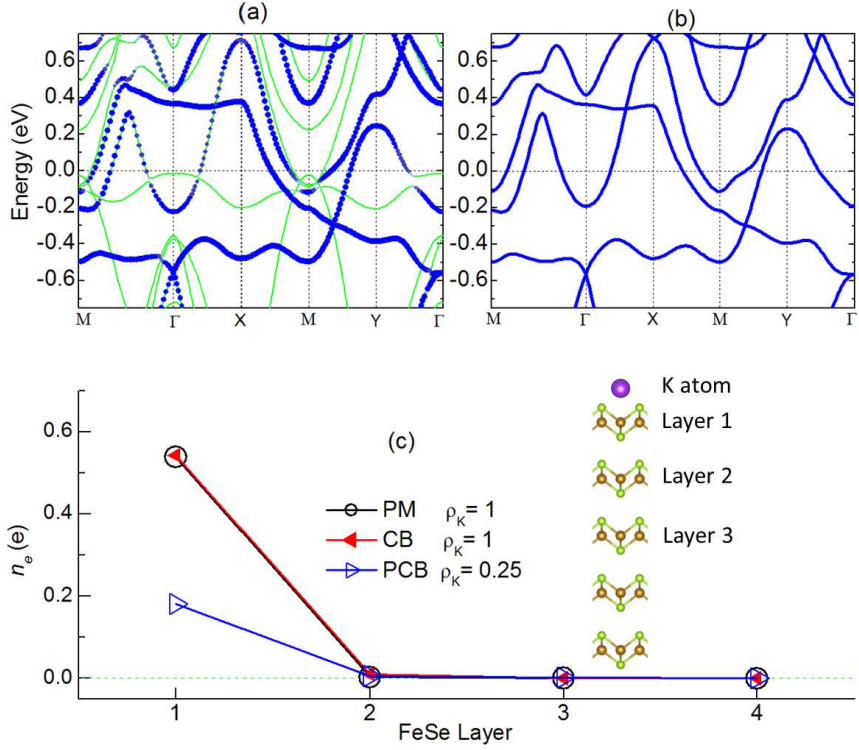}
\end{center}
\caption{(Color online) The energy band structures of K/2-FeSe/STO
(a) and K/FeSe (b) isolated systems. In panel (a), the energy bands projected
to K atom and top layer FeSe are signed by blue dots. The Fermi
level is set to be zero. Panel (c) shows the charge transfer from K atom to the top few FeSe layers in a K/n-FeSe system. The inner figure shows the K/n-FeSe atomic structure.}
\label{Fig2}%
\end{figure}

The energy bands of K/2-FeSe/SrTiO$_3$ with CB magnetic orientation are shown in Fig.
\ref{Fig2}(a). All the energy bands are plotted in green color, and
the energy bands that are mainly contributed by top layer FeSe and K atom are
signed by blue dots. To compare with these energy bands, we also
calculated the energy bands of K/FeSe system, which consists only of
the K atoms and top layer FeSe. The resultant band structure is
shown in Fig. \ref{Fig2}(b). It is quite similar to the blue dots in
Fig. \ref{Fig2}(a), while the energy bands in Fig. \ref{Fig2}(a)
without blue dots are quite similar to those in Fig. 2(a) in Ref.
\cite{zheng2013}. This means that the interaction between FeSe
layers is quite weak, and the atomic K adsorption only affects the
top layer FeSe. We then further checked the energy bands of
K/3-FeSe/SrTiO$_3$ system with CB magnetic orientation and K/2-FeSe/SrTiO$_3$ with PCB magnetic orientation.
The results are the same by the observed
fact that the adsorbed K atoms only influence the top layer FeSe and
have negligible effects on the other FeSe layers.
The doping charge distribution in a multilayer FeSe system can be
calculated by using a K/n-FeSe slab model. The calculation results of K/5-FeSe model with different $\rho_K$ and magnetic orientations are shown in Fig. \ref{Fig2}(c). All the calculation results show that the top layer FeSe has much more doping electrons than the second layer (more than 31 times larger). The doping electrons in the non-top layer FeSe is negligible. This phenomenon agrees with the most recent experimental results \cite{Seo2015}.
In other words,
the K/FeSe slab model can approximately describe the K adsorption on
n-FeSe/SrTiO$_3$ (n$\geq 2$) systems correctly. Thus, in this work,
we studied the detailed K adsorption effects by using K/FeSe slab
model with different values of $\rho_K$.

\begin{figure}[ptb]
\begin{center}
\includegraphics[width=0.7\linewidth]{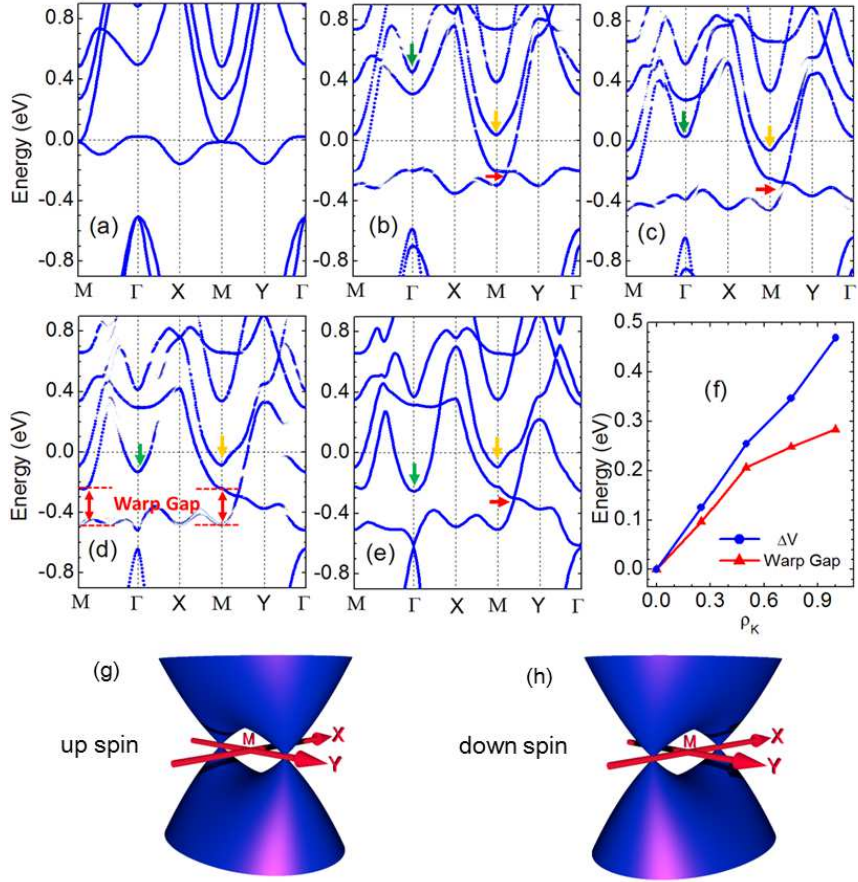}
\end{center}
\caption{(Color online) Spin-up unfolded electronic energy bands of
checkerboard-AFM FeSe monolayer with $\rho_K =$ 0.25 (b),  0.5 (c) and 0.75 (d). The
spin-up electronic energy bands of FeSe primary cell with (e) and
without (a) K atom adsorption are also shown here. Panel (f) shows
the warp gap at M point and electric potential difference $\Delta V$
as functions of $\rho_K$. Panels (g) and (h) show the double-Weyl cones around M point for up and down spin electrons respectively.}
\label{Fig3}%
\end{figure}

We then studied the energy bands of CB FeSe with different $\rho_K$.
The energy bands can be calculated by using a primary cell, if the
$\rho_K$ is 0 or 1. The corresponding results are shown in Figs.
\ref{Fig3}(a,e). For the cases of fractional $\rho_K$, supercells
have to be used to calculate their electronic structures. However,
the supercell energy bands are quite different from primary cell
energy bands. In order to compare energy bands between the primary
cell and the supercell, we resort to unfolding the supercell energy
bands by using a home-built code \cite{qu1}. In PAW calculations,
the norm of PAW wavefunction is not restricted to be unity.
Therefore, the energy band unfolding weight within PAW is not
accurate \cite{qu1}. Thus, we performed the QE calculations with
norm-conserving pseudopotentials to unfold the supercell energy
bands accurately. The plane wave cutoff energy was 1490 eV. The
$k$-space samplings were set the same with VASP calculations. We
tested that the FeSe and K/FeSe primary cell energy bands calculated
by QE agree well with VASP results. After the QE supercell
calculations, the {\it Wannier90} code \cite{wannier90} was used to
calculate the maximum localized Wannier functions (MLWFs). We added
a few Fortran lines to {\it Wannier90} to write out the MLWFs in a
format that is easier for further calculations.  After that, our
{\it Quantum Unfolding} code \cite{qu1,qu2} reads these MLWFs and
unfold the electronic energy bands to the first Brillouin zone of
FeSe primary cell.

We used 2$\times$2 CB FeSe supercells with 1, 2 and 3 K atoms adsorbed
on one side of FeSe monolayer to mimic the  K/n-FeSe/SrTiO$_3$
systems. The corresponding $\rho_K$ are 0.25, 0.5 and 0.75
respectively.  The unfolded energy bands are shown in Figs.
\ref{Fig3}(b-d). For comparison, the energy bands for the cases of
$\rho_K$=0 and $\rho_K$=1 (by usage of 1$\times$1 primary cells) are
also displayed in Fig. \ref{Fig3} [panels (a) and (e)].  Initially
[panel (a)], the Fermi surface contains two parts. They are a small
electron-like Fermi pocket at M point and a hole-like Fermi pocket
at $\Gamma$ point. With increasing $\rho_K$ step by step, then, the
energy bands continually go down with respect to the Fermi level. As
a consequence, the hole-like Fermi pocket disappears, while the
electron pocket enlarges its size dramatically. Remarkably, these
Fermi surface transformations shown in Figs. \ref{Fig3}(a-c) agree
well with the most recent ARPES experiments \cite{miyata2015}, in
which the K depositions with the electron doping up to 0.15 $e$ per
FeSe formula unit were explored. Prominently, as we increase
$\rho_K$, a conduction band (indicated by a green arrow) at $\Gamma$
point lowers its energy much faster than other energy bands. It
closes to the Fermi level at $\rho_K =$ 0.5, and crosses the Fermi
level at $\rho_K =$ 0.75. The crossing introduces an electron-like
Fermi pocket at $\Gamma$ point. This agrees with the previous energy
band calculation in uniformly electron doped FeSe monolayer
\cite{bazhirov2013}. Our unfolded energy bands show that this
phenomenon is robust against the doping charge distribution, and may
be realized in K/n-FeSe/SrTiO$_3$ system. It is a Lifshitz
transition since the Fermi surface topology is altered. The effect
of this Lifshitz transition to FeSe monolayer superconducting is
still not clear, therefore need more theoretical and experimental
study.

On the other hand, comparing to the case of uniformly electron doped
FeSe, the electron doping raised by atomic K adsorption brings about
a new feature. That is, the energy bands at M point become
non-symmetric, as shown by the red arrows in Fig. 3. They move to the
right side (M-Y direction). The bias direction depends on the
electron spin. We have checked that the spin-down energy bands bias
to the left side (M-X direction). This phenomenon also exists in
1-FeSe/SrTiO$_3$ system \cite{zheng2013}. Our DFT and tight-binding
calculations on FeSe monolayer in electric field revealed this
peculiar feature, and concluded that the energy bands biased at M
point is caused by a strong electric field along the perpendicular
direction of FeSe film. It is a self-established electric field as a
consequence of charge transfer. The strong electric field leads to
an electric potential difference between top and bottom Se atoms
($\Delta V\mathtt{=}V_{\text{top-Se}}\mathtt{-}V_{\text{bot-Se}}$).
The top (bottom) Se atoms are at the $\pm X$ ($\pm Y$) directions of
a Fe atom with spin-up magnetic moment, and are at $\pm Y$ ($\pm X$)
directions of a Fe atom with spin-down magnetic moment. Thus,
$\Delta V$ will lead to a difference between X and Y directions. We
studied the energy bands around the red arrow and find that they
compose a  double-Weyl cone structure.
Different from Dirac cone, the Weyl cone does not have spin degeneracy.
The sketches of  double-Weyl cones for
spin-up and spin-down electrons are shown in Figs. \ref{Fig3}(g,h).
The two Weyl points for up spin are in M-Y direction, while the two
Weyl points for down spin are in M-X direction. We name the energy
gap at M point as a warp gap, as shown in Fig. \ref{Fig3}(d). It
characterizes the warping strength of energy bands. The calculated
warp gap, as shown by the red line in Fig. \ref{Fig3}(f), is found
to linearly increase with the K coverage at the range of $0 < \rho_K
< 0.5$. Its slope decreases when  $\rho_K > 0.5$. We point out that
such kinds of warp gaps have been observed in the 1-FeSe/SrTiO$_3$
system \cite{cui2015}. The fitted yellow line in Ref. \cite{cui2015}
indicates the upper part of the double-Weyl cones in M-$\Gamma$
direction in Fig. 1(b) of Ref. \cite{cui2015}, wherein the bottom of
the yellow line clearly departs from the lower energy bands. The
estimated warp gap is about 50 meV.  The warp gap can also be
estimated from other ARPES experiments in 1-FeSe/SrTiO$_3$ system
\cite{he2013,liu2014}. From Figs. 2(b,c) in Ref. \cite{he2013} and
Fig. 2(d) in Ref. \cite{liu2014}, we estimated that the warp gap is
about 50 meV too. Based on our calculations, to observe the
double-Weyl cone energy band structure, the ARPES spectra should be
carried out along the M-X and M-Y directions. The ARPES spectrum
would show four Weyl cones around M point both along M-X and M-Y
directions, if the measurement does not distinguish the up and down
electron spins. From previous studies \cite{zheng2013}, we know that
the warp gap relates to $\Delta V$ directly. Thus, we calculated the
value of $\Delta V$ too. The electronic potential at a Se atom was
obtained by averaging the electronic potential at atom core area.
The calculated $\Delta V$ is shown by the blue line in Fig.
\ref{Fig3}(f). It linearly depends on $\rho_K$ in the whole K
coverage range we considered.

\begin{figure}[ptb]
\begin{center}
\includegraphics[width=0.7\linewidth]{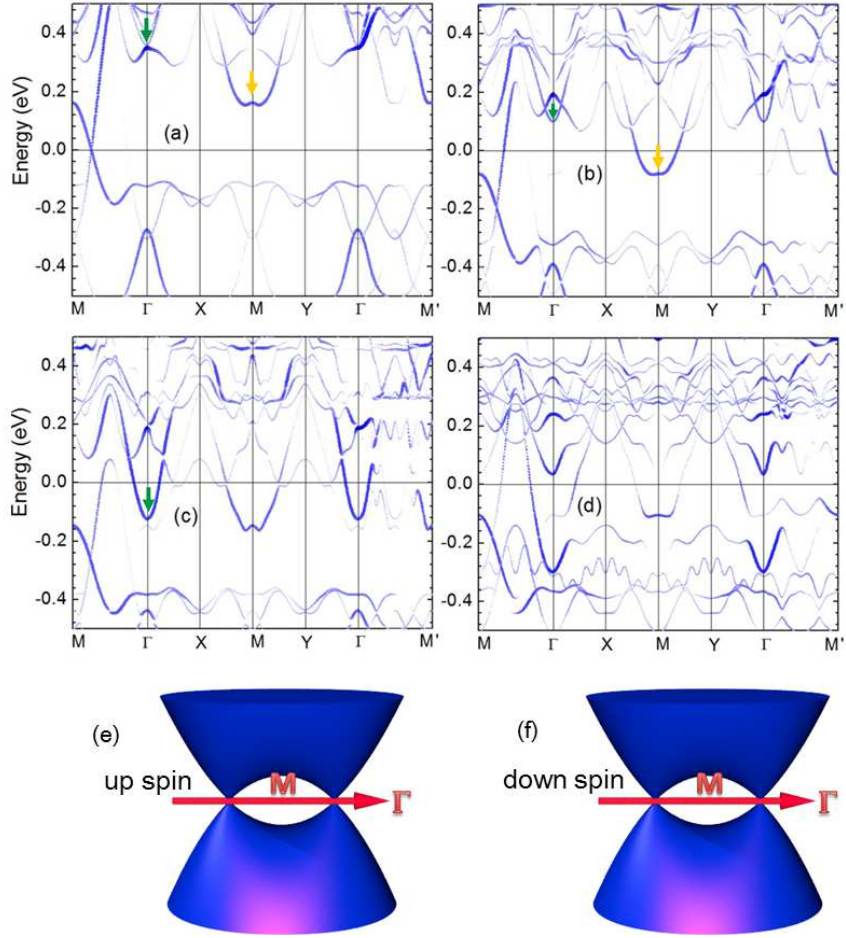}
\end{center}
\caption{(Color online) Spin-up unfolded electronic energy bands of
PCB FeSe monolayer with $\rho_K =$ 0 (a), 0.25 (b),  0.5 (c) and 0.75 (d). Panel (e) and (f) show the double-Dirac cones around M point for spin-up and spin-down electrons, respectively.}
\label{Fig4}%
\end{figure}

The unfolded PCB K/FeSe energy bands are shown in Figs. \ref{Fig4}(a-d).
For $\rho_K=0$ case, the unfolded energy bands in Fig. \ref{Fig4}(a) show that there are double-Dirac cones
around M point. It agrees with ARPES experiment results. Unlike the double-Weyl cones in CB FeSe, the double-Dirac cones in PCB FeSe
exist without doping electrons. The connecting line between the two Dirac points is in $\Gamma$-M direction.
They are the same for spin-up and spin-down electrons, as shown by the sketches in Figs. \ref{Fig4}(e,f).
The unfolding energy bands for $\rho_K=0$ show that there is no Fermi surface exists at $\Gamma$ point. The highest valence
energy bands locates at about 100 meV below Fermi energy. This conflicts with ARPES experiments, which show the
existence of hole-like Fermi pocket at $\Gamma$ piont. After K atom deposition ($\rho_K$=0.25), the whole
energy bands move downwards with respect to Fermi energy. The Dirac points locates about 200 meV below Fermi energy.
And the Fermi surfaces around M point combines to form a circle, as shown in Fig. \ref{Fig4}(b). The PCB FeSe Fermi surface
topology at $\rho_K$=0.25 is the same as that of CB FeSe. And it is consistent with ARPES experiment results. After we
increase K atom coverage to $\rho_K$=0.5, an electron-like Fermi pocket exists at $\Gamma$ point. It originates from the
energy band bottom signed by a green arrow in Figs. \ref{Fig4}(a-c). The energy band bottom crosses the Fermi level at 0.25$<\rho_K<$0.5.
The $\rho_K$ at the crossing point is smaller than that in CB FeSe, which is 0.5$<\rho_K<$0.75. As we further increase K atom
coverage to $\rho_K$=0.75, the energy band bottom at $\Gamma$ point lowers its energy, which increases the size of electron-like Fermi pocket.

\begin{figure}[ptb]
\begin{center}
\includegraphics[width=0.5\linewidth]{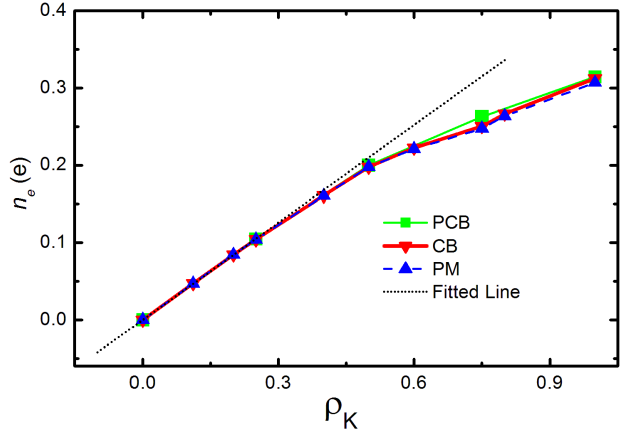}
\end{center}
\caption{(Color online) The charge doping ($n_e$) as a
function of K atom coverage ($\rho_K$) in paramagnetic (green blocks), checkerboard-AFM (red down triangles), and pair-checkerboard-AFM (blue up triangles) FeSe. The dotted black line is fitted from small $\rho_K$ data.}
\label{Fig5}%
\end{figure}

From the energy band variations, we know that atomic K adsorption
efficiently dopes electrons to the top FeSe layer. At this part, we
examined the amount of doped electrons by performing Bader analysis
\cite{bader}. In PCB FeSe, we used the $\sqrt{2}\times2\sqrt{2}$ supercell models with
0$\sim$4 K atoms. The calculation results are shown in Fig. \ref{Fig5} by green blocks. In CB FeSe, besides
primary cell model for FeSe and K/FeSe, we
also used three kinds of supercell models. They are 2$\times$2
supercell with 1$\sim$3 K atoms, $\sqrt{5}\times\sqrt{5}$ supercell
with 1$\sim$4 K atoms ($\rho_K =$ 0.2, 0.4, 0.6, 0.8), and
3$\times$3 supercell with 1 K atom ($\rho_K = \frac{1}{9}$). The
Bader analysis results are shown in Fig. \ref{Fig5} by red down triangles.
Besides the PCB and CB phases of FeSe monolayer, we also
studied the K adsorption on paramagnetic phase (PM) of FeSe
monolayer. The calculated electron transfer results are shown in
Fig. \ref{Fig5} by up triangles. The electron transfers for PCB, CB and PM
FeSe monolayers are the same for  $\rho_K <$ 0.5, and have a slight difference
for $\rho_K \geq$ 0.6. The maximum deviation is 0.016 $e$, locating at $\rho_K =$ 0.75.
Thus, we conclude that the magnetic state of FeSe has negligible effect on the
electron transfer from the adsorbed K atoms.
 The electron
doping increases with increasing $\rho_K$. It is linear when $\rho_K
<$ 0.5. We fitted calculated data points (in unit of $e$) at this
range by using ${n_{e}=c\times \rho_K}$.  The ${n_{e}}$ here is the
amount of doped electrons per FeSe formula unit. The fitted slope
${c}$ is 0.42. The value of $n_{e}$ deviates from the
${n_{e}=c\times \rho_K}$ line for $\rho_K >$ 0.5. It is lower than
the linear line value. The knee point locates between $\rho_K =$ 0.5
and 0.6. It is near the place where Lifshitz transition may happen in PCB and CB FeSe layer.

The reported maximum value of electron transfer in experiment
\cite{miyata2015} is 0.15 $e$. From the linear relation
${n_{e}=c\times \rho_K}$, we estimate that the $\rho_K$ is about
0.36. It is still in the linear range. The corresponding CB FeSe energy band
structure is in the interval between those shown in Figs.
\ref{Fig3}(b) and \ref{Fig3}(c), where the electron pocket at
$\Gamma$ point still does not emerge. The PCB FeSe energy bands for $\rho_K$ = 0.36
is in the interval between those shown in Figs. \ref{Fig4}(b) and \ref{Fig4}(c). It is the same interval where
Lifshiz transition occurs. However, our linear extrapolation show that Lifshiz transition in PCB FeSe occurs at
$\rho_K$ = 0.37, which is larger than 0.36. Thus, electron pocket at $\Gamma$ point does not emerge.
Both the calculation results at PCB and CB FeSe are consistent with the
ARPES experiment \cite{miyata2015}, which did not find any Fermi
surface around $\Gamma$ point. From the energy band structures in
Figs. \ref{Fig3}(a-e) and Figs. \ref{Fig4}(a-c), and the linear relation ${n_e=c\times
\rho_K}$, we estimate that the electron-like Fermi pocket at
$\Gamma$ point would appear for $n_e >$ 0.2 $e$ in CB FeSe and $n_e >$ 0.17 $e$ in PCB FeSe. To achieve this,
the simplest way is to deposit more K atoms on top layer FeSe. To
avoid the potential K-clustering problem sometimes occurred at large
coverages, here, we propose an alternative experiment that can dope
more electrons to FeSe monolayer. It is K deposition on monolayer
FeSe grown on SrTiO$_3$ (001) surface. After sufficient annealing,
SrTiO$_3$ substrate dope some amount of electrons to FeSe monolayer.
Then the K deposition on FeSe monolayer will further increase the
electron doping. The maximum $n_e$ is probably larger than 0.2 $e$.
Interestingly, a previous experiment has already found an electron
energy band centered at $\Gamma$ point, existing 75 meV above Fermi
energy \cite{huang2015}. It is very close to Fermi energy. The
energy bands in this experiment would lie in the interval of Figs.
\ref{Fig3}(b) and \ref{Fig3}(c), and is similar to Fig. \ref{Fig4}(b). Thus, we estimate that once we dope
more electrons to the system by K deposition, the
$\Gamma$ centered energy band would cross the Fermi level and
introduce an additional electron-like Fermi pocket.

In summary, we have theoretically studied the atomic K deposition on
n-FeSe/SrTiO$_3$ system. By using CB and PCB FeSe models, we explained
the Fermi surface topology observed in experiment, i.e., {the
existence of electron-like pockets at M point, as well as the disappearance
of the hole-like pockets and the enlargement of the electron-like
pockets after K deposition. Our results show that the K adsorbate
only affects the top layer FeSe. The charge transfer linearly
depends on the K coverage in the range $0 < \rho_K <$ 0.5 and leads
to a strong electric field along the perpendicular direction of FeSe
film.  This strong electric field warps the CB FeSe energy band structure at
M point and causes the occurrence of a double-Weyl cone structure. These
Weyl points locate at M-Y and M-X high-symmetry lines for spin-up
and spin-down electrons, respectively. Also, our energy band unfolding calculation shows that
PCB FeSe contain double-Dirac cones at Femi energy around M point for $\rho_K$ = 0. They are along $\Gamma$-M direction.
 We have also shown that an electron-like Fermi
pocket will emerge at $\Gamma$ point for $n_e >$ 0.2 $e$ in CB FeSe and $n_e >$ 0.17 $e$ in PCB FeSe, which is
expected to be observed in the undergoing atomic K deposition
experiments. Our obtained double-Weyl cones at M point and
electron-like pocket at $\Gamma$ point are typical features of
CB FeSe. The spin-resolved experiments in the future will figure out
whether CB or PCB magnetic orientation is the basement of the subsequent profound
physics for the FeSe monolayer and thus increase our knowledge about
this unique low-dimensional material.

This work was supported by Natural Science Foundation of China
(NSFC) under Grant Nos. 11474030 and 91321103, and by the National Basic Research Program of China (973 Program) under Grant
No. 2015CB921103.

\bibliographystyle{apsrev4-1}

\end{document}